\documentclass[letterpaper,twocolumn,english]{revtex4}
\usepackage[T1]{fontenc}
\usepackage[latin9]{inputenc}
\usepackage{xcolor}
\usepackage{amsthm}
\usepackage{amsmath}
\usepackage{amssymb}
\usepackage{graphicx}
\PassOptionsToPackage{normalem}{ulem}
\usepackage{ulem}

\makeatletter

\providecommand{\tabularnewline}{\\}
\providecolor{lyxadded}{rgb}{0,0,1}
\providecolor{lyxdeleted}{rgb}{1,0,0}

\@ifundefined{textcolor}{}
{%
 \definecolor{BLACK}{gray}{0}
 \definecolor{WHITE}{gray}{1}
 \definecolor{RED}{rgb}{1,0,0}
 \definecolor{GREEN}{rgb}{0,1,0}
 \definecolor{BLUE}{rgb}{0,0,1}
 \definecolor{CYAN}{cmyk}{1,0,0,0}
 \definecolor{MAGENTA}{cmyk}{0,1,0,0}
 \definecolor{YELLOW}{cmyk}{0,0,1,0}
}

\usepackage{babel}

\makeatother

\usepackage{babel}
\begin{document}

\title{The Swelling of Olympic Gels}

\author{M. Lang, J. Fischer, M. Werner, J.-U. Sommer}
\begin{abstract}
The swelling equilibrium of Olympic gels, which are composed of entangled
cyclic polymers, is studied by Monte Carlo Simulations. In contrast
to chemically crosslinked polymer networks, we observe that Olympic
gels made of chains with a \emph{larger} degree of polymerization,
$N$, exhibit a \emph{smaller} equilibrium swelling degree, $Q\propto N^{-0.28}\phi_{0}^{-0.72}$,
at the same polymer volume fraction $\phi_{0}$ at network preparation.
This observation is explained by a desinterspersion process of overlapping
non-concatenated rings upon swelling.
\end{abstract}
\maketitle
\begin{figure}
\begin{centering}
\includegraphics[width=0.5\columnwidth]{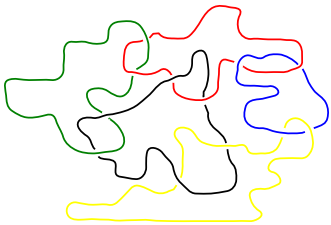}
\par\end{centering}

\caption{\label{fig:Sketch-of-the}Sketch of an Olympic gel.}
\end{figure}

Olympic gels \cite{key-28,key-3} are networks made of cyclic polymers
(``rings'') connected by the mutual topological inclusion of polymer
strands, see Fig. \ref{fig:Sketch-of-the}, with their elastic properties
depending exclusively on the degree of entanglements caused by the
linking of the rings. This particular difference to conventional polymer
networks and gels makes these materials an interesting model system,
since the pristine effect of entanglements on thermodynamic properties
of polymers is accessible. In particular, such gels could reveal the
role of entanglements for equilibrium swelling of polymer networks,
which is an outstanding problem in polymer physics. Since the term
Olympic gels has been coined by de Gennes \cite{key-28}, however,
the challenge to synthesize such materials has not been mastered yet,
although possible pathways for their synthesis have been proposed
\cite{key-3}.

\begin{figure}
\begin{centering}
\includegraphics[angle=270,width=0.9\columnwidth]{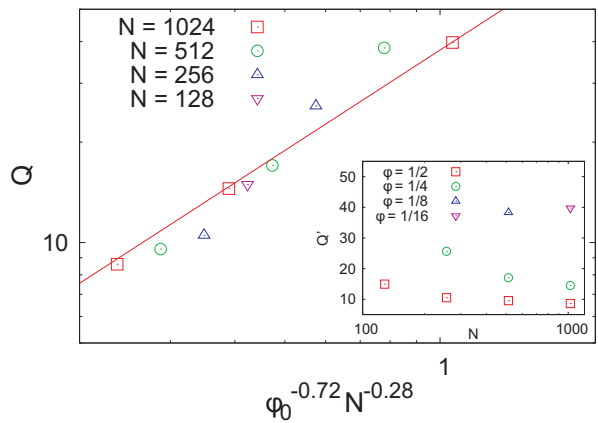}
\par\end{centering}

\caption{\label{fig:The-scaling-of}The scaling of the equilibrium degree of
swelling. Inset: unscaled data. The line indicates the proposed scaling
$Q\propto N^{-0.28}\phi_{0}^{-0.72}$. }
\end{figure}

In the present simulation study we construct Olympic gels, characterize
their topological state, and simulate isotropic swelling in athermal
solvent. We find that the equilibrium degree of swelling of Olympic
gels is described by a \emph{negative} power as function of the degree
of polymerization, $N$, of the rings, see Fig. \ref{fig:The-scaling-of},
in marked contrast to standard models of network swelling \cite{key-2}.
We will show that this result is a direct consequence of a desinterspersion
process originally proposed by Bastide \cite{key-17}, which allows
polymer rings to swell in part at no elastic deformation.

Let us first recall the essential predictions of the Flory-Rehner
(FR) \cite{key-1} model for equilibrium swelling of polymer networks.
The latter can be characterized by the equilibrium degree of swelling,
$Q$, which is defined by the ratio of the polymer volume at swelling
equilibrium with respect to the pure polymer volume in the dry state.
In the FR-model, it is assumed that the net change of free energy
upon swelling is solely given by the sum of the change of free energy
of mixing of the solvent with the polymer and the free energy change
of an affine elastic deformation of the network strands. One can express
this condition by equating the elastic osmotic pressure resulting
from an isotropic deformation of Gaussian chains and the osmotic pressure
of mixing, $\Pi_{el}=\Pi_{mix}$. The gel is prepared at a polymer
volume fraction $\phi_{0}$, where the elastically active network
chains (or rings) have an average extension of $R_{0}$. The equilibrium
degree of swelling of a polymer gel with an average elastic strand
length, $N$, is reached at a polymer volume fraction $\phi<\phi_{0}$,
for which network strands in solution exhibit an extension $R_{ref}$
solely due to excluded volume interactions. With these parameters,
the elastic ``pressure'' can be written as 
\begin{equation}
\Pi_{el}(\phi)\approx\frac{kT}{b^{3}}\frac{\phi}{N}\left(\frac{\lambda R_{0}}{R_{ref}}\right)^{2}\approx\frac{kT\phi}{Nb^{3}}\left(\frac{\phi_{0}}{\phi}\right)^{2/3}\left(\frac{\phi}{\phi_{0}}\right)^{\frac{2\nu-1}{3\nu-1}}\,\,.\label{eq:modulus}
\end{equation}
Here, $\lambda$ denotes the linear deformation ratio of the strands,
$\lambda^{3}=\phi_{0}/\phi$, $k$ is the Boltzmann constant, $T$
the absolute temperature, and $b$ denotes the root mean square length
of a Kuhn segment. The osmotic pressure due to mixing is given by
Des Cloizeaux's law \cite{key-28} 
\begin{equation}
\Pi_{mix}(\phi)\approx\frac{kT}{b^{3}}\phi^{3\nu/(3\nu-1)}\,\,.\label{eq:OsmPress}
\end{equation}
Both equations take into account the excluded volume effect for swelling
in good solvent with the exponent $\nu\approx0.588$ which is most
appropriate for the present simulation study. Equating both expressions
one obtains the equilibrium degree of swelling 
\begin{equation}
Q=\frac{1}{\phi}\approx N^{3(3\nu-1)/4}\phi_{0}^{-1/4}\approx N^{0.57}\phi_{0}^{-0.25}.\label{eq:Q}
\end{equation}
The hallmark of this text book result \cite{key-2} is that the equilibrium
degree of swelling \emph{grows} with increasing strand length $N$
and decreases weakly with increasing polymer volume fraction $\phi_{0}$
at preparation state. Eq.(\ref{eq:modulus}) is derived by assuming
an affine deformation of the chain ends and thus, it is assumed that
no swelling is possible without the free energy penalty of elastic
deformation. 

In Fig.~\ref{fig:The-scaling-of} we display our simulation results
for the equilibrium swelling of Olympic gels. In contrast to the FR
prediction we observe a \emph{reduction} of the degree of swelling
with increasing chain length, see inset of Fig.~\ref{fig:The-scaling-of}.
The best overlap of all simulation data is consistent with an ad hoc
scaling law given by 
\begin{equation}
Q\approx N^{-0.28}\phi_{0}^{-0.72}\label{eq:Q OG}
\end{equation}
as shown in the main plot of Figure \ref{fig:The-scaling-of}. We
will explain this unexpected behavior in this letter after a more
detailed analysis of the simulation data.

\begin{table}
\begin{centering}
\begin{tabular}{|c|c|c|c|c|c|c|c|c|c|}
\hline 
Sample  & \#1  & \#2  & \#3  & \#4  & \#5  & \#6  & \#7  & \#8  & \#9\tabularnewline
\hline 
\hline 
$N$  & 128  & 256  & 256  & 512  & 512  & 512  & 1024  & 1024  & 1024\tabularnewline
\hline 
$M$  & 1024  & 512  & 2048  & 2048  & 1024  & 512  & 1024  & 512  & 1024\tabularnewline
\hline 
$\phi_{0}$  & 0.5  & 0.5  & 0.25  & 0.5  & 0.25  & 0.125  & 0.5  & 0.25  & 0.0625\tabularnewline
\hline 
\hline 
$f_{n}$  & 2.7  & 5.64  & 2.89  & 10.9  & 6.04  & 2.98  & 17.7  & 10.08  & 2.76\tabularnewline
\hline 
$Q$  & 14.9  & 10.5  & 25.7  & 9.54  & 17.0  & 38.3  & 8.6  & 14.5  & 39.8\tabularnewline
\hline 
$D_{0}^{2}$  & 169.7  & 316  & 429  & 670  & 855  & 1056  & 1252  & 1581  & 2442\tabularnewline
\hline 
$D^{2}$  & 513  & 1191 & 1140  & 2618  & 2726  & 2608 & 5354 & 5790 & 5310\tabularnewline
\hline 
$P_{0}$  & 2.70  & 4.50  & 2.83  & 6.28  & 4.08  & 2.44  & 10.3  & 7.10  & 2.89\tabularnewline
\hline 
$P$  & 1.15  & 2.80  & 1.10  & 5.81  & 3.12  & 1.21  & 9.8  & 5.96  & 1.32\tabularnewline
\hline 
\end{tabular}
\par\end{centering}

\caption{\label{tab:-is-the}$N$ is the degree of polymerization of the rings,
$M$ the number of rings per sample, $\phi_{0}$ the polymer volume
fraction at preparation conditions, $f_{n}$ the average number of
concatenated pairs of rings per ring, $Q$ the equilibrium degree
of swelling, $D^{2}$ and $D_{0}^{2}$ are the square average distances
of two opposite monomers of a ring in the swollen and the preparation
state, $P$ and $P_{0}$ are the overlap numbers in the swollen state
and the preparation state.\textbf{ }}
\end{table}

To simulate Olympic gels we used a GPU-version \cite{key-14} of the
bond fluctuation method \cite{key-7}, which is an efficient simulation
method for polymers in the semi-dilute and concentrated regime \cite{key-111}.
The preparation of the samples is identical to our previous work \cite{key-6}
for the concatenated series of melts except of using a non-periodic
box as simulation container. By including diagonal moves in the preparation
step of Olympic gels we allow the crossing of bonds without a change
in the excluded volume constraints. By returning to the original set
of moves the thus created topology is conserved. The key parameters
of the samples are summarized in table \ref{tab:-is-the}. In the
present work, we focus on samples with an average number of concatenations
per ring $f_{n}\ge2$, for which we can identify a well developed
dominant largest cluster (gel). $f_{n}$ is determined as described
in \cite{key-6} and follows the prediction 
\begin{equation}
f_{n}\propto\phi_{0}^{\nu/(3\nu-1)}N\propto\phi_{0}^{0.77}N.\label{eq:fn-1}
\end{equation}
After preparation, the networks are placed into the middle of a large
simulation container and swollen to equilibrium, which was monitored
by the drop in the polymer volume fraction near the middle of the
gel. Empty lattice sites model a perfect athermal solvent. The equilibrium
degree of swelling $Q$ is determined by analyzing $\phi^{-1}$ for
the innermost 50\% of the monomers. We consider $\phi=0.5$ as melt
concentration with the reference value $Q=1$. The polymer volume
fraction at swelling equilibrium for any sample is below 1/16, which
justifies a semi-dilute approximation of chain conformations. The
overlap number of a given ring, $P$, is determined by counting the
centers of mass of other rings in a sphere with radius $D$ around
the center of mass of each ring, whereby $D$ is the average distance
of two opposite monomers of a ring. For convenience, we also use $D$
to measure the deformation of the rings. Note that the chains are
only weakly deformed with a maximum $D/D_{0}\approx2.07$ for all
samples.

\begin{figure}
\begin{centering}
\includegraphics[angle=270,width=1\columnwidth]{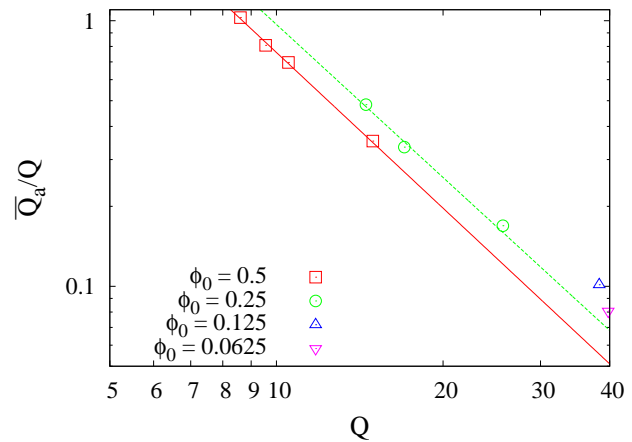}
\par\end{centering}

\caption{\label{fig:Fraction-of-affine}The fraction of the apparent affine
contribution $\overline{Q}_{a}$ to the equilibrium degree of swelling
$Q$. The lines indicate best fits with power laws $\propto Q^{-1.9\pm0.2}$
and $\propto Q^{-1.95\pm0.08}$ for $\phi_{0}=1/4$ and $\phi_{\text{0}}=1/2$
respectively.}
\end{figure}

In Figure \ref{fig:Fraction-of-affine}, we display the apparent affine
deformation part $\overline{Q}_{a}/Q$ of swelling given by 
\begin{equation}
\overline{Q}_{a}=\left(D/D_{0}\right)^{3},\label{eq:Qa}
\end{equation}
where $D_{0}$ is the ring extension at preparation conditions. According
to the FR-modell we have $Q\equiv\overline{Q}_{a}$ per definition
\cite{key-2,key-31}. Figure \ref{fig:Fraction-of-affine} displays
$\bar{Q}$ as a function of the degree of equilibrium swelling. Large
values for $Q$ are obtained by a non-affine swelling while the limit
of small $Q$ is well described by an apparently fully affine deformation
of the chains. The data of Figure \ref{fig:Fraction-of-affine} indicate
a relation in the vicinity of $\overline{Q}_{a}/Q\propto Q^{-1.95}$
for all samples of our study with a small additional correction as
function of $\phi_{0}$. 

\begin{figure}
\begin{centering}
\includegraphics[angle=270,width=1\columnwidth]{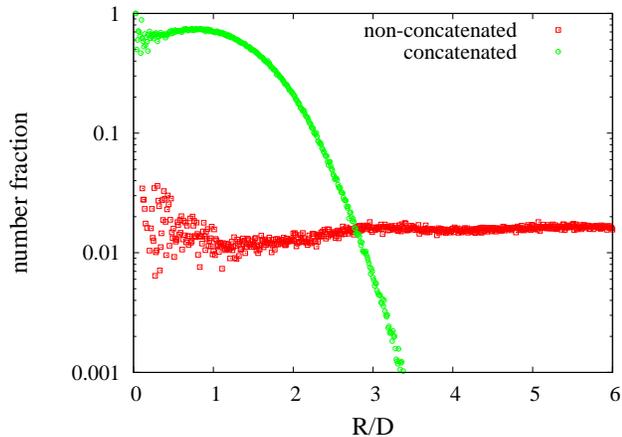}
\par\end{centering}

\caption{\label{fig:Distance-distribution-of-1}Normalized distance distribution
between centers of mass of previously overlapping non-concatenated
and concatenated rings at swelling equilibrium (sample \#6).}
\end{figure}

One possible mechanism for the observed non-affine swelling is the
rearrangement of cyclic polymers upon swelling without elastic deformation.
To identify such rearrangements, we distinguish between concatenated
and non-concatenated rings which are overlapping \emph{at preparation}
\emph{conditions}. The distance distributions of centers of mass of
these ring populations is then analyzed \emph{at swelling equilibrium}.
The data of the two overlapping populations of sample \#6, see Tab.\ref{tab:-is-the},
with small $f_{n}$ is shown in Figure \ref{fig:Distance-distribution-of-1}
as an example. The data show that non-concatenated rings essentially
are squeezed out of the volume $4\pi D^{3}/3$ while the concatenated
rings remain within a distance of order $D$.

\begin{figure}
\begin{centering}
\includegraphics[angle=270,width=1\columnwidth]{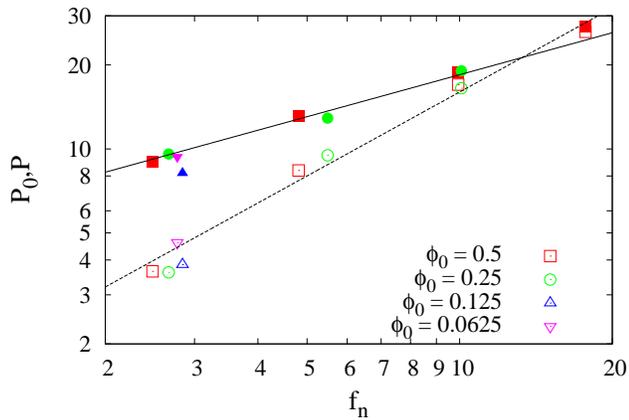}
\par\end{centering}

\caption{\label{fig:Disinterpenetration-of-rings}The overlap number of cyclic
polymers in the preparation state, $P_{0}$ (filled symbols), at swelling
equilibrium, $P$ (hollow symbols) as function of the average number
of concatenations per ring, $f_{n}$. The dashed line indicates $P\propto f_{n}$
and the solid line indicates $P_{0}\propto f_{n}^{1/2}$.}
\end{figure}

As consequence, the overlap number, $P$, at swelling equilibrium
is roughly proportional to $f_{n}$ for all samples of our study as
shown in Figure \ref{fig:Disinterpenetration-of-rings}. Note that
$f_{n}$ grows linearly with $N$ and thus, more rapidly than the
Flory number $P$ of overlapping molecules. As mentioned in Ref. \cite{key-6},
$f_{n}$ must converge towards $P_{0}$ for large $N$. Convergence
is nearly reached for the samples with the largest values of $f_{n}$.
To show this convergence, we added the data for $P_{0}\propto f_{n}\phi_{0}^{0.27}\propto\phi_{0}^{0.65}N^{1/2}$
in Fig. \ref{fig:Disinterpenetration-of-rings} ignoring the weak
extra $\phi_{0}$ - dependence of $P_{0}$.

Based upon the above observations, we argue that\emph{ }the dominating
contribution to the non-affine swelling stems from the \emph{desinterspersion}
\emph{of non-concatenated rings} upon swelling in the partially concatenated
regime with $f_{n}\propto N$. To derive the equilibrium swelling
condition in this regime, we assume full desinterspersion of overlapping
non-concatenated rings\emph{ }and an affine deformation of the concatenated
rings\emph{.} For the sake of the argument let us introduce an intermediate
state of swelling that we call the ``desinterspersed state'' and
denote this state by subscript $'des'$. In the desinterspersed state,
the total number of correlation volumes per volume of a ring, $R_{des}^{2}/\xi_{des}^{3}$,
can be approximated by the number of blobs per chain, $N/g_{des}$
times the number $f_{n}\propto\phi_{0}^{\nu/(3\nu-1)}N$ of overlapping
concatenated chains 
\begin{equation}
\frac{R_{des}^{3}}{\xi_{des}^{3}}\approx\frac{N}{g_{des}}f_{n}.\label{eq:blobs_number}
\end{equation}
We consider the polymer volume fraction at the desinterspersed state
\begin{equation}
\phi_{des}\approx\frac{b^{3}g_{des}}{\xi_{des}^{3}}\propto\frac{b^{3}N^{2}}{R_{des}^{3}}\phi_{0}^{\nu/(3\nu-1)}\label{eq:deformation_onset}
\end{equation}
as reference state for the onset of the affine deformation. The size
of a non-deformed ring 
\begin{equation}
R_{des}\approx bN^{1/2}\phi_{des}^{-(\nu-1/2)/(3\nu-1)}\label{eq:Initial_size}
\end{equation}
at polymer volume fraction $\phi_{des}$ leads to a degree of swelling
in the desinterspersed state 
\begin{equation}
Q_{des}=1/\phi_{des}\sim N^{-(3\nu-1)}\phi_{0}^{-2\nu}\sim N^{-0.76}\phi_{0}^{-1.18}.\label{eq:deformation_onset_II}
\end{equation}
This result is the key to understand the negative power for $N$ at
the equilibrium degree of swelling. 

For $f_{n}>1$, desinterspersion must stop at a polymer volume fraction
larger than the overlap concentration $\phi^{*}\propto N^{-(3\nu-1)}$
proposed by De Gennes \cite{key-28}. In fact, we find $\phi_{des}\propto1/\phi^{*}$,
which shows that desinterspersion becomes increasingly difficult with
increasing overlap of the rings. Since $\phi_{des}\gg\phi^{*}$, swelling
equilibrium is reached by an additional elastic deformation of the
rings. We consider only the permanent entanglements as approximated
by the number of concatenations to be relevant at swelling equilibrium
and assume that higher topological invariants are not important for
the partially concatenated regime with $f_{n}\propto N$. To apply
the affine model for deformation, we subdivide the $N$ segments of
the ring into $f_{n}$ elastic chains by assuming that for small $f_{n}\lesssim10$
all concatenated chains are deforming the concatenating ring at swelling
equilibrium. Swelling equilibrium is found by using $\phi_{des}$
as new ``preparation condition'' instead of $\phi_{0}$ in Eq. (\ref{eq:modulus}).
This leads to 
\begin{equation}
Q\approx\left(\frac{N}{f_{n}(\phi_{0})}\right)^{3(3\nu-1)/4}\phi_{des}^{-1/4}\approx N^{-(3\nu-1)/4}\phi_{0}^{-5\nu/4}\label{eq:Q new}
\end{equation}
and thus, $Q\approx N^{-0.19}\phi_{0}^{-0.74}$, which is in good
agreement with our ad hoc scaling prediction in Eq.(\ref{eq:Q OG})
for the simulation data.

As direct consequence of this model we find that the apparent affine
fraction of swelling depends on the desinterspersed state

\begin{equation}
\overline{Q}_{a}=\left(\frac{R_{g}}{R_{g,0}}\right)^{3}=\left(\frac{R_{des}}{R_{g,0}}\right)^{3}\left(\frac{R_{g}}{R_{des}}\right)^{3}=\left(\frac{R_{des}}{R_{g,0}}\right)^{3}Q_{a}.\label{eq:Qa-1}
\end{equation}
Since the true affine fraction $Q_{a}/Q=1/Q_{des}$ is related to
the equilibrium degree of swelling by $Q_{a}(Q)/Q\propto Q_{des}^{-1}(Q)\propto Q^{-4}$,
see Eq.(\ref{eq:Q new}), the apparent affine fraction of swelling
is also universal, i.e. is independent of the length of the rings:
\begin{equation}
\frac{\overline{Q}_{a}}{Q}=\left(\frac{R_{des}}{R_{g,0}}\right)^{3}\frac{Q_{a}}{Q}\propto\phi_{0}^{3(\nu-1/2)/(3\nu-1)}Q^{-2/(3\nu-1)}\label{eq:Qad}
\end{equation}
with a strong dependence, $\overline{Q}_{a}/Q\propto Q^{-2.62}$,
on $Q$ . A similar \emph{universality} is observed in Fig.~\ref{fig:Fraction-of-affine},
which is a striking evidence for the existence of the desinterspersion
process. 

To conclude, Olympic gels display a highly non-affine swelling behavior
due to desinterspersion processes, if the linking number $f_{n}$
is smaller than the Flory number $P$. This latter condition characterizes
the partially concatenated regime, for which a pairwise analysis of
linked states seems to be sufficient \cite{key-6}. The good qualitative
agreement between simulation data and model further indicates that
each concatenation may contribute a pair of elastic strands to the
network, which might be a reasonable approximation for the partially
concatenated regime. 

It is important to point out here, that the structure of \emph{any}
network can be decomposed into a set of connected cycles \cite{key-29},
whereby the average cycle size is of order 8 chains for typical strand
lengths around 50-100 Kuhn segments between 4-functional junctions
\cite{key-24}. Therefore, most elastomers are located in the regime
$f_{n}\propto N$ where desinterspersion of non-concatenated cyclic
structures upon swelling occurs. Based upon our results, therefore,
we expect a clear impact of desinterspersion onto the equilibirum
swelling degree of polymer gels. This view is supported by simulations
that detect a non-affine swelling of crosslinked networks on length
scales much larger than the size of individual network strands \cite{key-20}
and by experiments that measure a vanishing non-affine contribution
to elasticity at large degrees of swelling \cite{key-30}. Scattering
and NMR data indicate that the initial swelling may be dominated by
a desinterspersion process that is followed by a deformation of the
chains \cite{key-17}, which is in full accord with our model but
in opposite order as the assumption used by Painter and Shenoy \cite{key-21}.
In particular, the length scale at which the deformation becomes affine
is an essential parameter that needs to be understood in the framework
of elasticity models that predict a non-affine deformation behavior
\cite{key-31}. Our analysis of Olympic gels gives a fresh view on
the problem of swelling of polymer gels in general and reveals that
connectivity caused by topological concatenation can lead to a qualitativly
different swelling behavior.

ML thanks T. Kreer and A. Galuschko for stimulating discussions and
the DFG for funding grant LA2735/2-1.

\end{document}